\documentclass[12pt]{article}

\title{Multi-criteria decision making via multivariate quantiles}

\author{
Daniel Kostner\footnote{Free University Bozen, Faculty of Economics and Management, \href{mailto:daniel.kostner@economics.unibz.it}{daniel.kostner@unibz.it}}
}
\date{{\small \today}}

\usepackage{array} 
\usepackage{graphicx}
\usepackage{float} 
\usepackage{caption, subcaption} 
\usepackage[margin=1cm,font=small,labelfont=bf]{caption}
\usepackage{amsmath, amssymb, enumerate, mathrsfs, color}
\usepackage[colorlinks=true, linkcolor=blue, citecolor=blue, urlcolor=blue, filecolor=blue]{hyperref} 

\newtheorem{theorem}{Theorem}

\newtheorem{definition}[theorem]{Definition}

\newtheorem{example}[theorem]{Example}

\numberwithin{equation}{section}  
\numberwithin{figure}{section}    
\numberwithin{table}{section}     
\numberwithin{theorem}{section}

\newcommand{\of}[1]{\ensuremath{\left( #1 \right)}}

\newcommand{\cb}[1]{\ensuremath{ \left\{ #1 \right\} }}

\newcommand{\bs}{\backslash}

\newcommand{\R}{\mathrm{I\negthinspace R}}

\newcommand{\Int}{{\rm int\,}}

\begin{document}

\maketitle

\abstract{A novel approach for solving a multiple judge, multiple criteria decision making (MCDM) problem is proposed. The ranking of alternatives that are evaluated based on multiple criteria is difficult, since the presence of multiple criteria leads to a non-total order relation. This issue is handled by reinterpreting the MCDM problem as a multivariate statistics one and by applying the concepts in \cite{HamelKostner18JMVA}. A function that ranks alternatives as well as additional functions that categorize alternatives into sets of ``good'' and ``bad'' choices are presented. Moreover, the paper shows that the properties of these functions ensure a logical and reasonable decision making process.}


\section{Introduction}
The aim of this work is to propose a new method for solving multiple criteria decision making (MCDM) problems. The basic idea is to interpret the MCDM problem as one of multivariate statistics and apply new set optimization methods. More specifically, a modified version of the multiple judge, multiple criteria ranking problem in \cite{Buckley1984} is solved by replacing the fuzzy set approach by a new one based on set-valued quantiles for multivariate random variables. This actually is a new way to deal with the fundamental difficulty--the lack of a natural complete order for multidimensional objects--in MCDM. 

The paper \cite{Buckley1984} investigates a multiple criteria decision problem which employs the testimony of judges to make an informed choice. The judges (experts) supply both the information on how the alternatives satisfy each criterion and the information about the importance of the criteria. This information is provided in form of two fuzzy sets. Fuzzy set theory --introduced in \cite{Zadeh1965} and \cite{Klaua1965}-- is an extension of classical set theory; fuzzy sets are sets whose elements have degrees of membership. In classical set theory, the membership of elements in a set is assessed in binary terms (an element either belongs or it does not). Fuzzy set theory permits the gradual assessment of the membership of elements in a set, by means of a membership function with values in the real interval $[0,1]$. This enables to solve many problems that deal with imprecise and uncertain data by allowing for imprecise input.

In the present paper, we propose an alternative approach to a slightly modified problem, where the information on how the alternatives satisfy each criterion is provided by a third party. The $C$-quantiles from \cite{HamelKostner18JMVA} perfectly fit the requirements for solving this problem: the judges' opinions on the importance of the criteria are modeled via a vector order generated by a cone $C$ and the alternatives with their criteria ratings are interpreted as realizations of a random vector. It turns out that for a sufficiently large parameter (probability) $p$ the $C$-quantiles extract the alternatives that best match the judges' specifications.

Howard Raiffa and Robert Schlaifer with \cite{SchlaiferRaiffa1961} as well as of Ron Howard with \cite{Howard1968} had an important influence on modern decision theory. Nowadays, MCDM is a sub-discipline of operations research that explicitly evaluates multiple conflicting criteria in decision making. As well described by \cite{Grabisch2016}: ``\textit{Decision under multiple criteria deals with situations where an agent (called the decision maker) has to choose between several objects, alternatives, options,etc. (called hereafter alternatives), considering together several points of view or criteria pertaining to different aspects, descriptors or attributes, which describe the alternatives under consideration.}'' The difficulty of this decision problem is that, ``\textit{there are antagonistic points of view: some alternatives may be best preferred under some point of view, but are much less attractive under another point of view. The fundamental difficulty behind is simply that there is no natural complete order on multidimensional objects.}'' (\cite{Grabisch2016}). The last quote has many counterparts in the statistical literature since the lack of a ``canonical'' quantile (function) in the multivariate case usually is attributed to this ``lack of a natural order in higher dimensions'', e.g., \cite[p. 1126]{BelloniWinkler11AS}. The concepts introduced in \cite{HamelKostner18JMVA} deal with this problem. Moreover, these new notions enjoy basically all the properties of their univariate counterpart and are based on the recent developments in set optimization as surveyed in \cite{HamelEtAl15Incoll}.

The paper is structured as follows. In the first section, the ranking problem of \cite{Buckley1984} is presented and a modified version is adapted to accomodate the new concepts. In the second section, the cone distribution function and the set-valued quantiles from \cite{HamelKostner18JMVA} are applied to the decision problem. The properties of these functions and their use in the decision making process are analyzed. Simple examples and figures are provided as illustrations.

\section{The multiple judge, multiple criteria ranking problem}
\label{sec:buck}

The decision making problem discussed in \cite{Buckley1984} is the following. The aim is to rank a set of alternatives $A = \{a_1, a_2,...,a_m\}$ from ``best'' to ``worst'', based on a set of criteria $\Gamma = \{\gamma_1, \gamma_2,...,\gamma_d\}$. This is accomplished by employing judges (advisors, experts) $\mathscr{J} = \{J_1, J_2,...,J_n \}$. The judges use a scale $\mathscr{L}=\{S_0, S_1,...,S_L \}$ of preference information to assess the criteria's importance and the criteria of the alternatives. It is assumed that $\mathscr{L}$ is linearly ordered: $S_1<S_2 \cdots <S_L$. No other structure is assumed to exist on $\mathscr{L}$. Only ordinal information is required from the judges, usually $S_i$ are not numbers, hence $\mathscr{L}$ is an ordinal scale.

\begin{example}
\label{ex:ordinalscale}
$\mathscr{L}=\{\emptyset, VL, L, M, H, VH, P \}$, where $\emptyset=$none, $VL=$very low, $L=$low, $M=$medium, $H=$high, $VH=$very high, and $P=$perfect.
\end{example}

Each judge $J_j$ indicates the relative importance (weight) of criterion $\gamma_k$, $v_j : \Gamma \to \mathscr{L}$. Moreover, each judge $J_j$ supplies the information of how alternative $a_i$ satisfies criterion $\gamma_k$, $x_j : \Gamma \times A \to \mathscr{L}$. The same scale $\mathscr{L}$ is used for $x_j(\gamma_k,a_i)$ and $v_j(\gamma_k)$, hence $x_j(\gamma_k,a_i), v_j(\gamma_k) \in \mathscr{L}$. Therefore, each judge $J_j$ has two sets:
\[
X^j = \begin{bmatrix}
   	x_j(\gamma_1,a_1) & \dots  & x_j(\gamma_1,a_m) \\
	\vdots & \ddots & \vdots \\
	x_j(\gamma_d,a_1) & \dots  & x_j(\gamma_d,a_m)
\end{bmatrix} \in \mathscr{L}^{d \times m}
\]
and
\[
v^j = \begin{bmatrix}
   	v_j(\gamma_1) \\
	\vdots \\
	v_j(\gamma_d)
\end{bmatrix} \in \mathscr{L}^d.
\]

In \cite{Buckley1984}, the sets above are denoted as fuzzy sets, even if their membership functions have values in $\mathscr{L}$ (instead of $[0,1]$). Three possible applications for the problem outlined above are described. First, an agency consults experts for awarding grants. Second, an environmental bureau relies on scientists to rank certain chemicals from most harmful to least harmful. Third, a government agency ranks, with the help of high ranking officials, energy sources based on different criteria. For more details on the examples please refer to \cite{Buckley1984}. Moreover, the paper \cite{Buckley1984} discusses methods of aggregating all the sets to achieve a set of ranked alternatives. More specifically, it examines when and how to pool the judges' information as well as how to compute the final ranking.

The aim of this work is to propose a new method to solve a decision making problem, which is similar to the one discussed in \cite{Buckley1984}. The MCDM problem is translated into the context of multivariate statistics and solved by set optimization methods with tools from \cite{HamelKostner18JMVA}. Compare \cite{HamelEtAl15Incoll} for details about the complete lattice approach to set optimization, which is the basis for \cite{HamelKostner18JMVA}. The difference between the MCDM in \cite{Buckley1984} and the one solved in this paper is that the evaluations $x_j(\gamma_k,a_i)$ are provided by a third party. This novel approach to MCDM has the benefit that it enables both an analytic and a geometrical representation of the problem.

\section{A set optimization approach to MCDM}
\label{sec:setvalued}

Our problem is as follows. As in \cite{Buckley1984}, the goal is to rank a set of alternatives $A = \{a_1, a_2,...,a_m\}$, taking into account a set of criteria $\Gamma = \{ \gamma_1, \gamma_2,...,\gamma_d \}$. However, the information is taken from two different sources. On the one hand, each judge $J_1, J_2,...,J_n$ indicates the relative importance of criterion $\gamma_k$ with respect to the other criteria via a positive number, $v_j : \Gamma \to \R_+$. This preference information is based on a ratio scale $\mathscr{S}$. Nevertheless, the judges can also provide ordinal information as in example \ref{ex:ordinalscale}, as long as it can be translated into a ratio scale.

\begin{example}
\label{ex:scale}
If $\mathscr{S} = \{1,2,3\}$, then the judge has the possibility to assign to every criterion $\gamma_k$ the integers 1, 2 or 3. Hence, the relative importance of a criterion with respect to another criterion can be expressed in seven possible weights: $\frac{1}{1}, \frac{1}{2}, \frac{1}{3}, \frac{2}{1}, \frac{2}{3}, \frac{3}{1}, \frac{3}{2}$.
\end{example}

Therefore, each judge has a basket of weights:
\[
v^j = \begin{bmatrix}
   	v_j(\gamma_1) \\
	\vdots \\
	v_j(\gamma_d)
\end{bmatrix} \in \R^d_+,
\]
called the \textit{importance vector}. 
On the other hand, the information of how $a_i$ satisfies criterion $\gamma_k$ is provided by an external evaluator/source, $x_E : \Gamma \times A \to \R$. This enables the customization of MCDM to applications in finance, where the external source could be the asset market. Each alternative $a_i$ is assessed via the vector
\[
x^E(a_i) = \begin{bmatrix}
   	x_E(\gamma_1,a_i) \\
	\vdots \\
	x_E(\gamma_d,a_i),
\end{bmatrix} \in \R^d
\]
whereas $X$ stores the evaluation of all alternatives:
\[
X =	\begin{bmatrix}
	   	x^E(a_1) & \dots  & x^E(a_m)
	\end{bmatrix} 
	= 
	\begin{bmatrix}
   		x_E(\gamma_1,a_1) & \dots  & x_E(\gamma_1,a_m) \\
		\vdots & \ddots & \vdots \\
		x_E(\gamma_d,a_1) & \dots  & x_E(\gamma_d,a_m)
	\end{bmatrix} 
	\in \R^{d \times m}.
\]
Below, $X$ will be understood as realizations of a $d$-dimensional random vector.

Let us recall a few useful concepts from the theory of ordered vector spaces. A preorder is a reflexive and transitive relation, and an antisymmetric preorder is called a partial order. The set $H^+(w) = \cb{z \in \R^d \mid w^\top z \geq 0}$ is called the closed homogeneous halfspace with normal $w$. A set $C \in \R^d$ is called a closed convex cone if $s > 0, z \in C$ imply $sz \in C$ and it is closed under addition, i.e., $x,y  \in C$ implies $x+y \in C$. The vector preorder $\leq_C$ is generated by the convex cone $C \in \R^d$ with $0 \in C$ by means of
\[
x \leq_C y \Leftrightarrow y - x \in C. 
\]

\subsection{Modeling the judges' preferences}
\label{subsec:cones}

The data presented in the section above is used to introduce two convex cones. These cones generate a vector order which is crucial for the ranking of the alternatives.

The $n$ importance vectors $v^j$ of the judges generate a cone via 
\[
K_I = \cb{\sum_{j=1}^n s_jv^j \mid s_1,\hdots,s_n \geq 0}.
\] 
This convex cone pools the judges opinion on the relative importance of each criteria. It is called the \textit{importance cone} $K_I$. If some of the importance vectors $v^j$ are non-negative linear combinations of others, already the latter ones span $K_I$. These $v^j$ can be interpreted as the ``extreme judges''. In general, the cone includes all non-negative linear combinations of these ``extreme'' vectors. From an application standpoint, the inclusion of all non-negative linear combinations means that not only the judges views are incorporated, but also all potential compromises between the judges. 

The halfspace that has the weight vector $v^j$ as normal
\[
 H^+(v^j) = \cb{z \in \R^d \mid {v^j}^\top z \geq 0}
\] 
can be interpreted as the acceptance set of the judge $j$. The halfspace $H^+(v^j)$ generates the total preorder
\[
x \leq_{ H^+(v^j)}  y,
\]
which ranks the criteria evaluations of the alternatives--$x^E(a_i)$ depending on the judge's importance vector $v^j$. By taking the intersection over such halfspaces with the $n$ vectors $v^j$ as normals another polyhedral convex cone can be formed: 
\[
K_A = \bigcap_{j=1}^n  H^+(v^j).
\]
This cone incorporates the view of all judges. It represents the positions with respect to the criteria (evaluations) that are unanimously accepted by all judges. It is called the \textit{acceptance cone} $K_A$ and it generates a vector preorder on the set $X$:
\begin{equation*}
x \leq_{K_A} y.
\end{equation*}

In summary, $K_A$ represents the accepted positions in the criteria and $K_I$ the relative importance given to each criteria.

The approach regarding the cones is taken differently in this work compared to the one in \cite{HamelKostner18JMVA}. On the one hand, the goal in \cite{HamelKostner18JMVA} is to incorporate a vector preorder $\leq_C$ given by a cone $C$ into the statistical analysis of multivariate data. The dual cone of $C$ comes into play from a technical standpoint, as $\leq_C$ can be represented as intersection of total preorders generated by the closed halfspaces $H^+\of{w}$ for $w \in C^+\bs\{0\}$. On the other hand, in this paper the judges' opinions on the relevance of each criterion $v^j$ constitute the cone $K_I$. However, the order relation that ranks the criteria assessments of the alternatives $x^E(a_i)$ is given by the acceptance cone $K_A$.

The cone $K_A$ resembles the solvency cone, an important concept in financial mathematics, see \cite{HamelHeyde2010}. $K_I$ is a finitely generated convex cone and hence closed. Since $v^j \in \R^d_+$ it follows that  $K_I \subseteq \R^d_+$. Moreover, due to the bipolar theorem it holds that: 
\[
K_I = (K_A)^+ = \cb{w \in \R^d \mid  \forall z \in K_A \colon z^\top w \geq 0},
\]
and vice versa. Consequently, the importance cone $K_I$ and the acceptance cone $K_A$ are dual to each other. This is due to the fact that $K_I$ is spanned by the importance vectors $(v^1,\hdots,v^n)$, which are orthogonal to the acceptance sets of the judges. To summarize the discussion, we give an abstract definition.

\begin{definition}
An importance cone $K_I$ is any closed convex cone such that $\{0\} \subsetneq K_I \subseteq \R^d_+$, whereas an acceptance cone $K_A$ is any closed convex cone such that $\R^d_+ \subseteq K_A \subsetneq \R^d$.
\label{def:acceptanceimportancecone}
\end{definition}

\begin{example}
In order to illustrate the intuition behind these new concepts, a decision making example with two criteria is used. Figure \ref{fig:judgeHs} shows a two-dimensional importance vector $v^1 = (2,1)$, where the criterion on the horizontal axis is twice as relevant as the criterion on the vertical axis. Moreover it depicts a closed homogeneous halfspace $H^+(v^1)$ (blue area) with normal $v^1$, defined as $H^+(v^1) = \cb{z \in \R^d \mid {v^1}^\top z \geq 0}$. The judge is indifferent between the criteria evaluations of the alternatives--$x^E(a_i)$ on the boundary of this halfspace as well as between those on a line that is parallel to the boundary and included in the halfspace. However, the judge is not indifferent between the assessed alternatives--$x^E(a_i)$ on different lines. Therefore, $H^+(v^1)$ can be seen as the acceptance set of the judge.

\begin{figure}[H]
\centering
\includegraphics[width=0.4\textwidth]{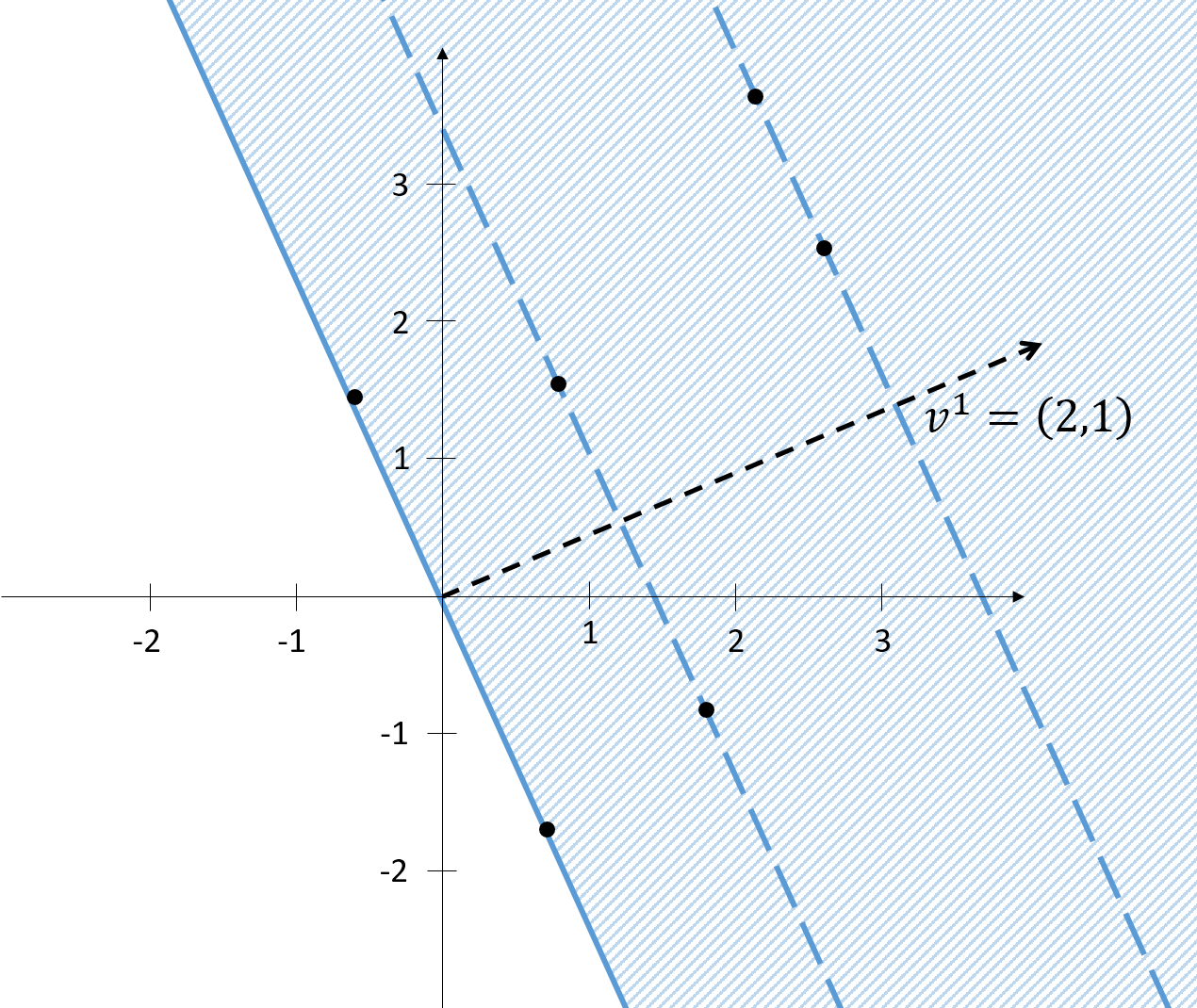}
\caption{The importance vector $v^1$ and its acceptance set $H^+(v^1)$.}
\label{fig:judgeHs}
\end{figure}

Now, if there are several judges all their opinions should be incorporated. Figure \ref{fig:figureCones}a shows the value given (dotted vectors) to the criteria by three judges: $v^1 = (2,1)$, $v^2 = (1,1)$ and $v^3 = (1,2)$. The first and the third judge deem one criteria more important than the other, whereas the second assigns equal significance to the criteria. Again, these vectors are normals to the acceptance sets (blue halfspaces) of the judges. In order to derive an acceptance set that complies to all judges, the intersection (\ref{fig:figureCones}b) over all their acceptance sets is taken. From this follows the acceptance cone $K_A$ (light yellow cone in \ref{fig:figureCones}b) that represents the positions in the criteria accepted by all judges. As illustrated in figure \ref{fig:figureCones}c the importance cone (yellow opaque) is only spanned by $v^1 = (2,1)$ and $v^3 = (1,2)$, as $v^2 = (1,1)$ is a convex combination of the former two.
 
\begin{figure}[H]
\centering
\begin{minipage}{.3\textwidth}
  \centering
  \includegraphics[width=0.95\textwidth]{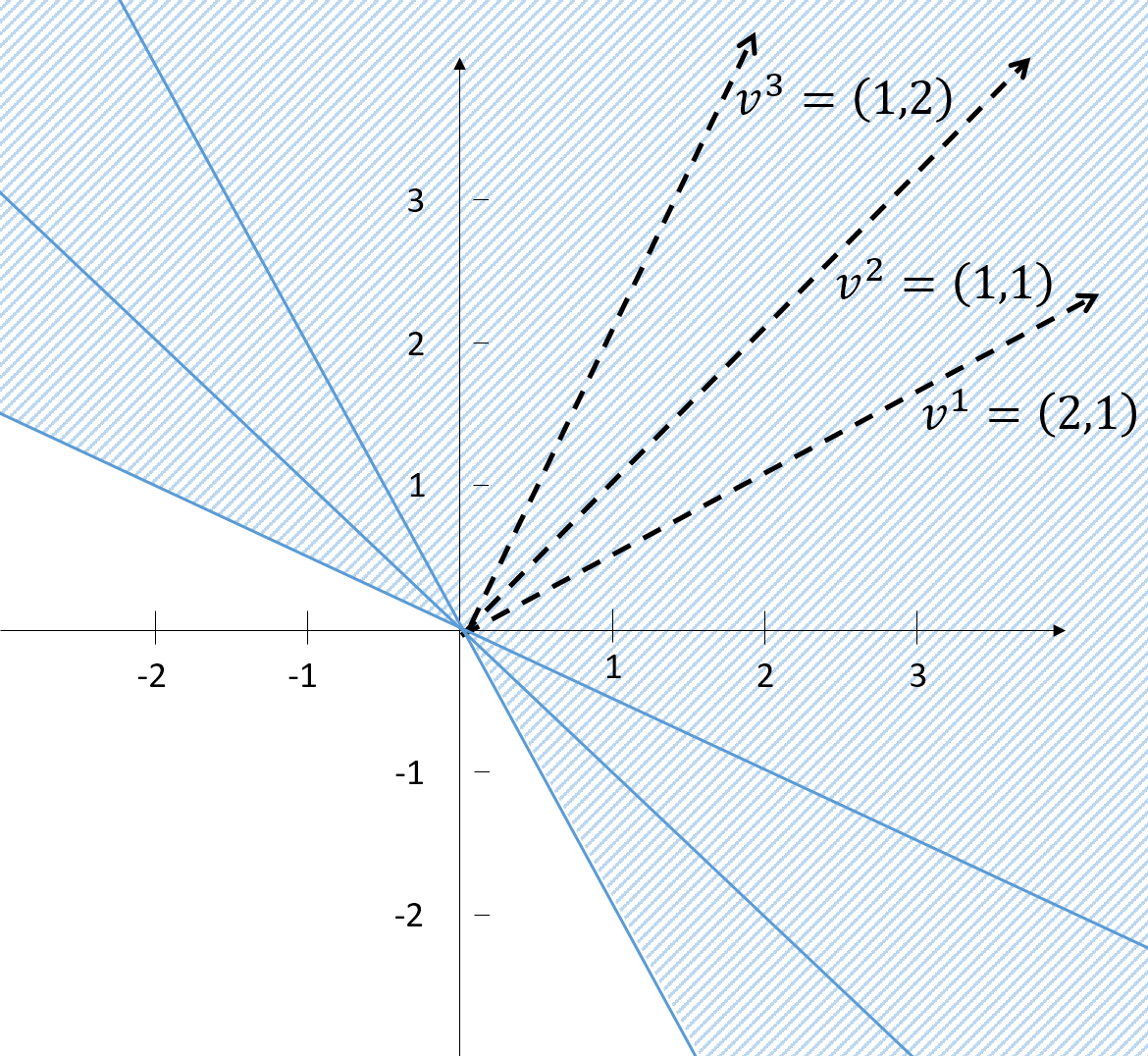}
  \caption*{(a)}
  \label{fig:judgesCones}
\end{minipage}
\begin{minipage}{.3\textwidth}
  \centering
  \includegraphics[width=0.95\textwidth]{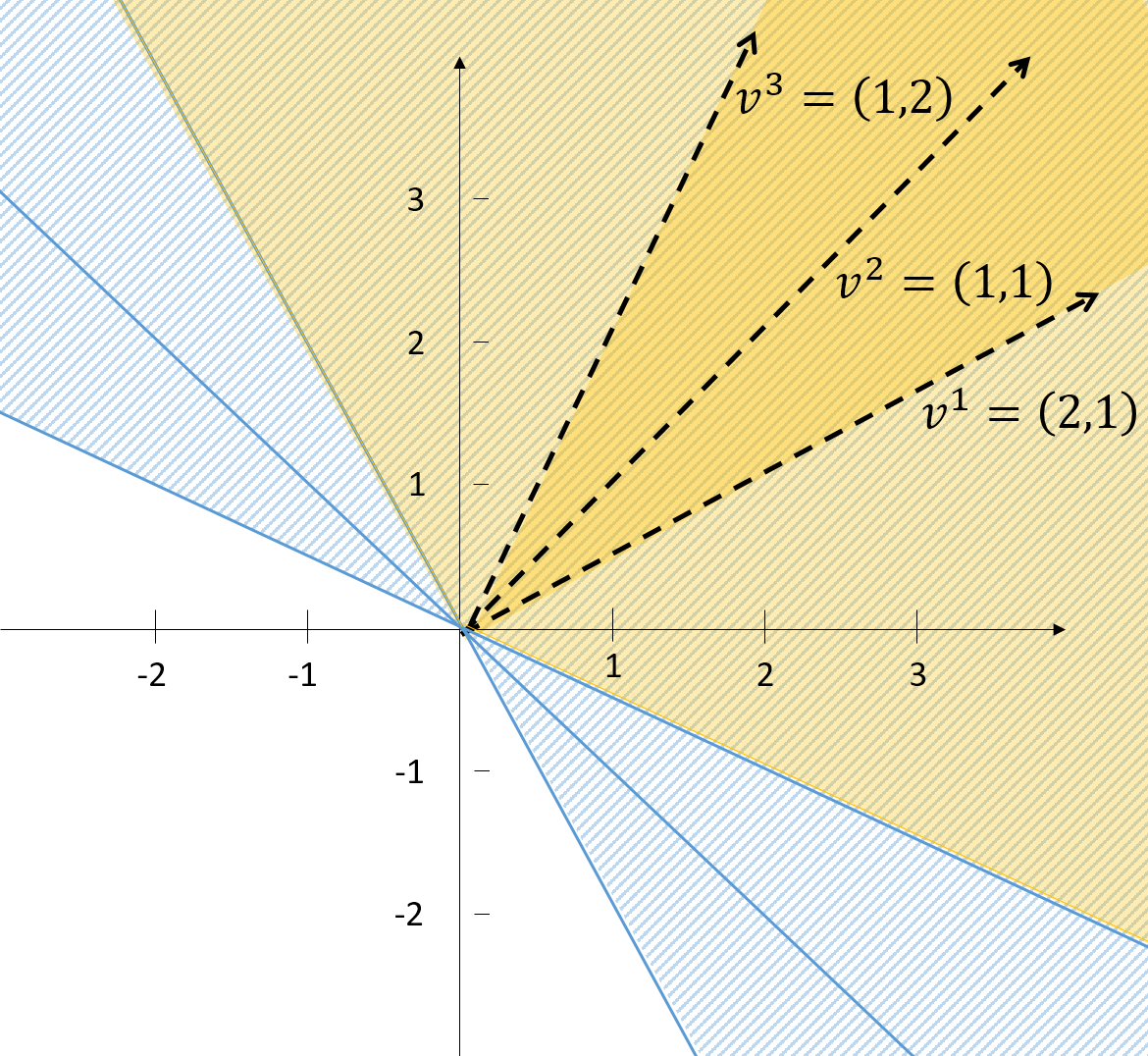}
  \caption*{(b)}
  \label{fig:judgesConesIntersection}
\end{minipage}
\begin{minipage}{.3\textwidth}
  \centering
  \includegraphics[width=0.95\textwidth]{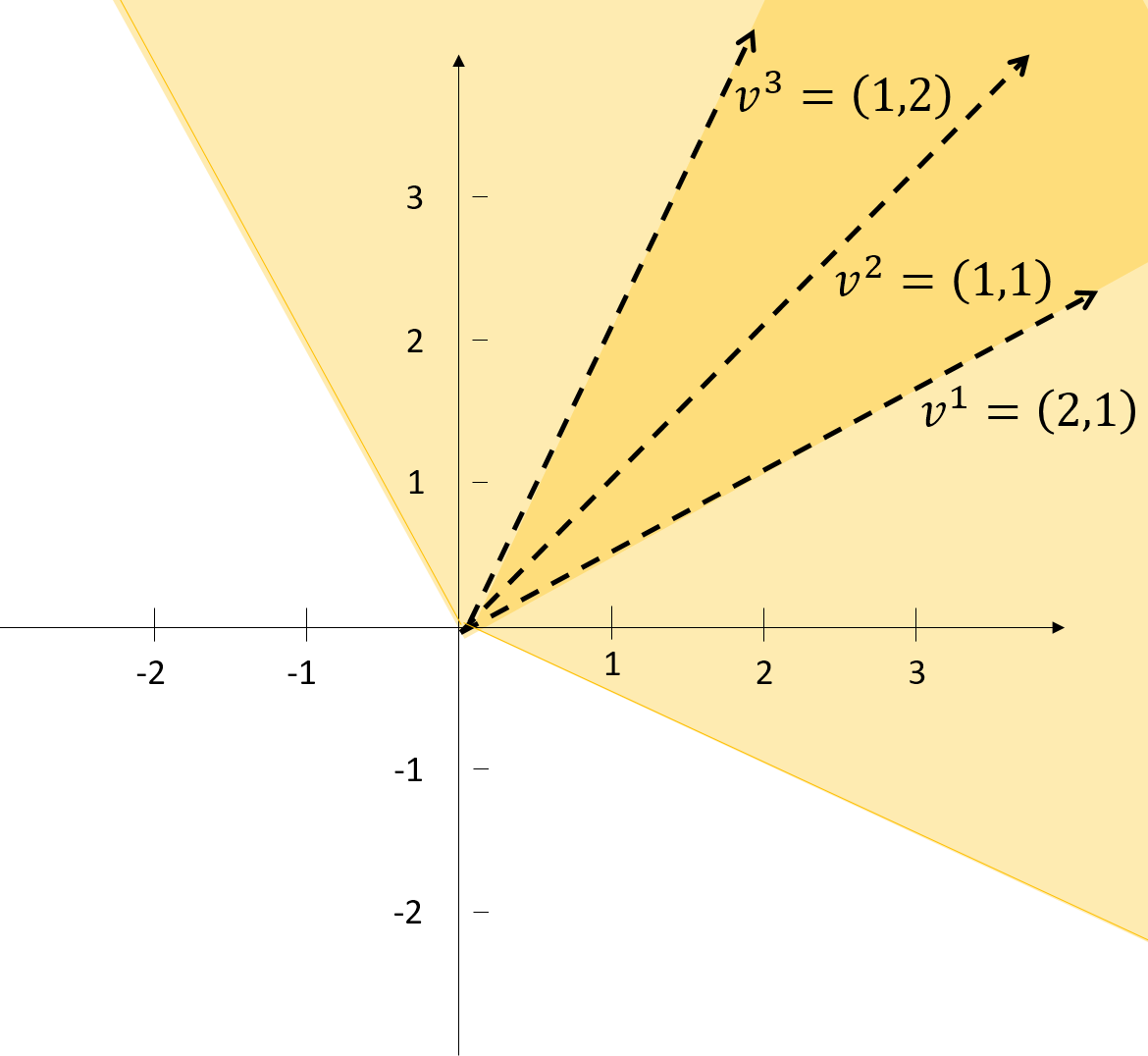}
  \caption*{(c)}
  \label{fig:cones12}
\end{minipage}
\caption{From the individual importance vectors and acceptance sets to the cones $K_I$ and $K_A$.}
\label{fig:figureCones}
\end{figure}
\end{example}

\begin{figure}[H]
\centering
\includegraphics[width=0.4\textwidth]{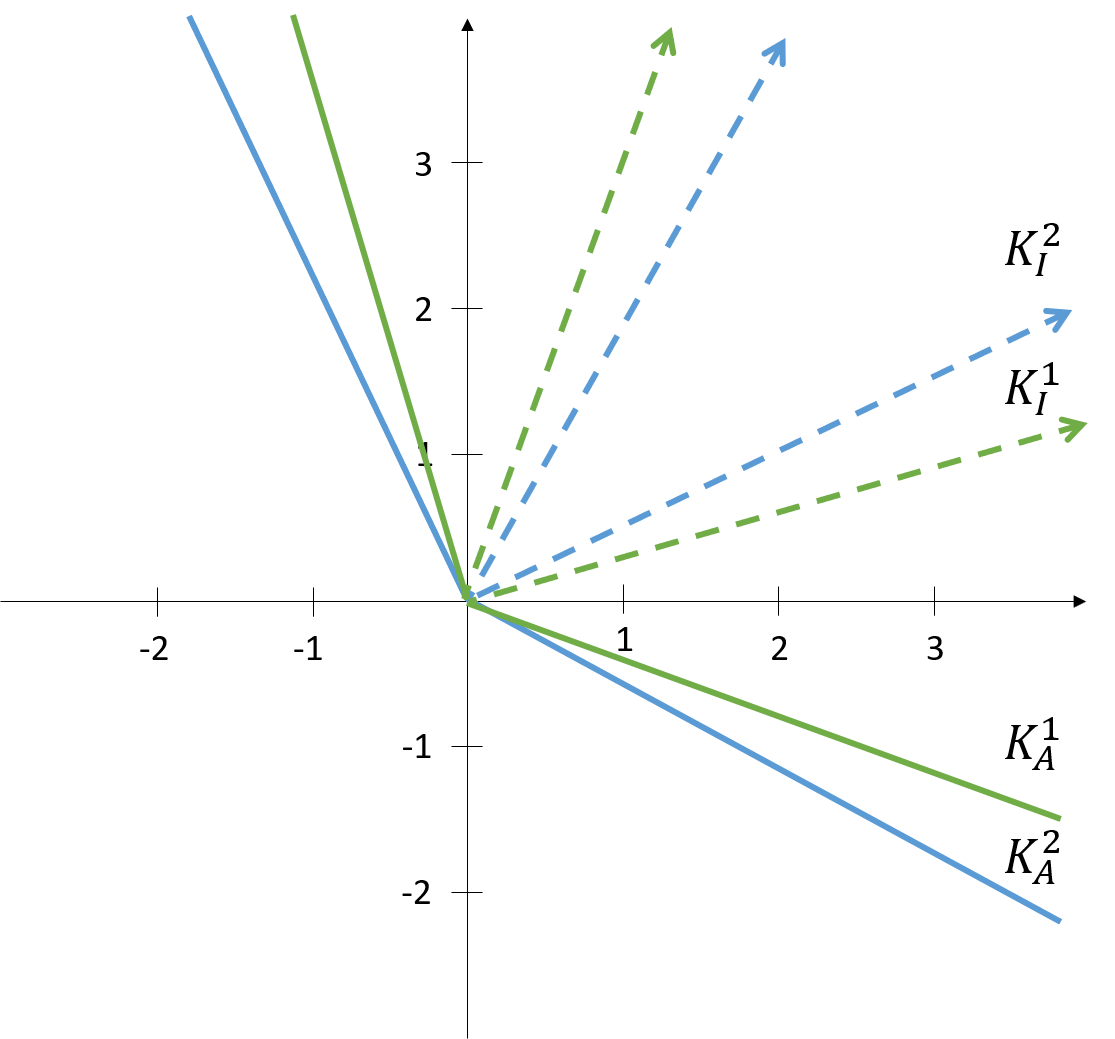}
\caption{The interplay between the judges' opinions and the cones.}
\label{fig:ConeInCone}
\end{figure}

\begin{example}
An interesting illustration on how the cones reflect the judges views is shown in figure \ref{fig:ConeInCone}. The more the ``extreme'' judges differ from each other, the wider the importance cone $K^1_I \supseteq K^2_I$ and the smaller the acceptance cone $K^1_A \subseteq K^2_A$. Further differing opinions on the criteria imply that a greater compromise has to be made, which takes shape in a smaller acceptance cone $K_A$.
\end{example}

\subsection{Cone distribution functions}
\label{subsec:conedistribution}

In this section, the MCDM problem is translated into a multivariate statistics framework. The set of alternatives $A=\{a_1,a_2,\hdots,a_m\}$ is understood as the sample set $\Omega$ of a vector-valued random variable $X$ with uniform distribution. This implies that each $x^E(a_i) \in X$ has the same probability. More precisely, $(\Omega, \mathcal{F}, \Pr)$ is the probability space with $\Omega = A$, $\mathcal{F} = \mathcal P(A)$ and $\Pr$ is the uniform probability measure (distribution), whereas $X : \Omega \to \R^d$ is a $d$-dimensional random variable.

The recently introduced cone distribution function (see \cite{HamelKostner18JMVA}) is used in order to rank the alternatives.  Therefore, it is reinterpreted as a ranking function of the elements in $X$, which associates to a $x^E(a_i) \in X$ a rank $p$ from $0$ to $1$. From now on the parameter $p$ is interpreted as a rank (order) indicator and denoted as rank. The cone distribution function of $X$ is defined in a two step procedure. 

First, fix $v \in K_I\bs\{0\}$ and consider $F_{X,v}(z) \colon \R^d \to [0,1]$ defined by
\[
F_{X,v}(z) = F_{v^\top X}(v^\top z) = \Pr\of{v^\top X \leq v^\top z},
\] 
where the random vector $X$ and the function's input $z$ is scalarized by the importance vector $v$. In fact, $v^\top X$ is a random variable and $F_{X,v}(z)$ is its cumulative distribution function taken at $v^\top z$. This distribution function can be reformulated as the probability of $X$ being in the halfspace $ - H^+(v) = \cb{z \in \R^d \mid v^\top z \leq 0}$ with normal $v$ and $z$ on its boundary:
\[
F_{X,v}(z) = \Pr\of{v^\top X \leq v^\top z} = \Pr\of{X \in z - H^+(v)}.
\]
If $X = \{x_E(a_1),\hdots,x_E(a_m)\}$ is the set of the criteria evaluations of all alternatives and $z=x_E(a)$ is the criteria assessments of the alternative $a$, then $F_{X,v}(z)$ quantifies how much $x_E(a_i) \in X$ are ranked below $x_E(a)$ by a judge with the importance vector $v$. In fact, the halfspace $z - H^+(v)$ is the acceptance set directed downwards and with the criteria evaluations of alternative $a$ on its boundary.

Since the scalar distribution function is applied on a finite sample set, it makes sense to define its empirical version:
\[
\tilde F_{X,v}(z) = \frac{1}{m} \#\cb{x_E(a_i) \mid a_i \in A, x_E(a_i) \in z - H^+(v)}.
\]

Second, all the judges' importance vectors are considered jointly to rank the criteria evaluations of the alternatives $x^E(a_i) \in X$. This can be done by taking the infimum over all elements of the importance cone $K_I$. The use of $K_I$ implies that not only the judges' importance vectors $v$ are considerd but also their non-negative linear combinations. This has the beneficial effect that the optimal solution does not have to compel to a single judge, but can also be a compromise between judges. It follows that for every $v \in K_I$ the vector $v$ which assigns to $x^E(a)$ the lowest rank via $\tilde F_{X,v}(x^E(a))$ is chosen. Hence, a conservative/risk-limiting ranking of the assessed alternatives is ensured. This is exactly what the cone distribution function $F_{X,C} \colon \R^d \to [0,1]$ in definition 2.1 of \cite{HamelKostner18JMVA} with $C=K_A $ does:
\[
F_{X, K_A}(z) = \inf_{w \in K_I\bs\{0\}} F_{X,w}(z) =\inf_{w \in K_I\bs\{0\}} \Pr\of{X \in z - H^+(w)}.
\]
The cone distribution function takes the acceptance cone as input, since it ranks the $x^E(a_i)$ based on the vector preorder $\leq_{K_A}$. This derives from the fact that the relation $\leq_{K_A}$ can be represented by a family of scalar functions, i.e.,
\[
z \leq_{K_A} y \quad \Leftrightarrow \quad \forall v \in K_I \colon v^\top z \leq v^\top y \quad \Leftrightarrow \quad \forall v \in K_I \colon z \leq_{H^+(v)}y.
\]
This means that $\leq_{K_A}$ is represented as intersection of the acceptance sets $H^+(v)$, for all $v \in K_I\bs\{0\}$. 

The empirical variant of the cone distribution function adapted to the MCDM is as follows:
\[
\tilde F_{X,K_A}(z) = \inf_{v \in K_I\bs\{0\}} \frac{1}{m} \#\cb{x_E(a_i) \mid a_i \in A, x_E(a_i) \in z - H^+(v)}.
\] 

The properties of the function $\tilde F_{X,K_A}(z)$ (see \cite{HamelKostner18JMVA}) which have the most interesting meaning for the multiple criteria decision making problem are reviewed.

\textbf{a)} {\em Affine equivariance}, i.e., if $b \in \R^d$ and $A \in \R^{d \times d}$ is an invertible matrix, then
\[ 
\forall z \in \R^d \colon \tilde F_{AX + b, AK_A}\of{Az + b} = \tilde F_{X, K_A}\of{z}.
\]
On one side, if the scale of the criteria has a proportional variation for all criteria ($b$), then the ranking of the assessed alternatives $x^E(a_i)$ is not compromised. On the other side, if only some criteria undergo a change in scale ($A$), then this must be reflected in the acceptance cone, respectively in the importance cone.

\textbf{b)} {\em Monotone non-decreasing} function of $z$ with respect to $\leq_{K_A}$, i.e., if $y \leq_{K_A} z$, then $\tilde F_{X, K_A}\of{y} \leq \tilde F_{X, K_A}\of{z}$. This property states that $\tilde F_{X, K_A}$ ranks with respect to the preorder given by the acceptance cone.

\textbf{c)} {\em Monotone non-increasing} function of $X$ with respect to $\leq_{K_A}$, i.e., if $X \leq_{K_A} Y$, then $\tilde F_{X, K_A}\of{z} \geq \tilde F_{Y, K_A}\of{z}$ for all $z \in \R^d$. The rank of an element depends on how the element compares against the others. If there are two sets of evaluated alternatives $X$ and $Y$, where $Y$ includes better rated elements with respect to the judges tastes, then an element included in both sets gets a lower ranking with the set including the better elements $Y$.  

\textbf{d)} If $\emptyset \neq K_A^1 \subseteq K_A^2 \subset \R^d$ are two closed convex cones, then $\tilde F_{X, K_A^1}(z) \leq F_{X, K_A^2}(z)$ for all $z \in \R^d$. The judges' views have an important effect on the ranking. As illustrated in figure \ref{fig:ConeInCone}, the more apart the opinions of the judges on the criteria are, the wider is the importance cone, consequently the narrower is the acceptance cone -- with the latter cone implying a bigger compromise to be made. This effort of accommodating all the judges' opinions is reflected by the cone distribution function in assigning for the same alternative a lower rank when the acceptance cone gets narrower.

From the discussion above, it is clear that the cone distribution function $\tilde F_{X,K_A}(z)$ is perfectly capable of ordering a set of alternatives by integrating the opinions of multiple judges.

\begin{figure}[H]
\centering
\includegraphics[width=0.4\textwidth]{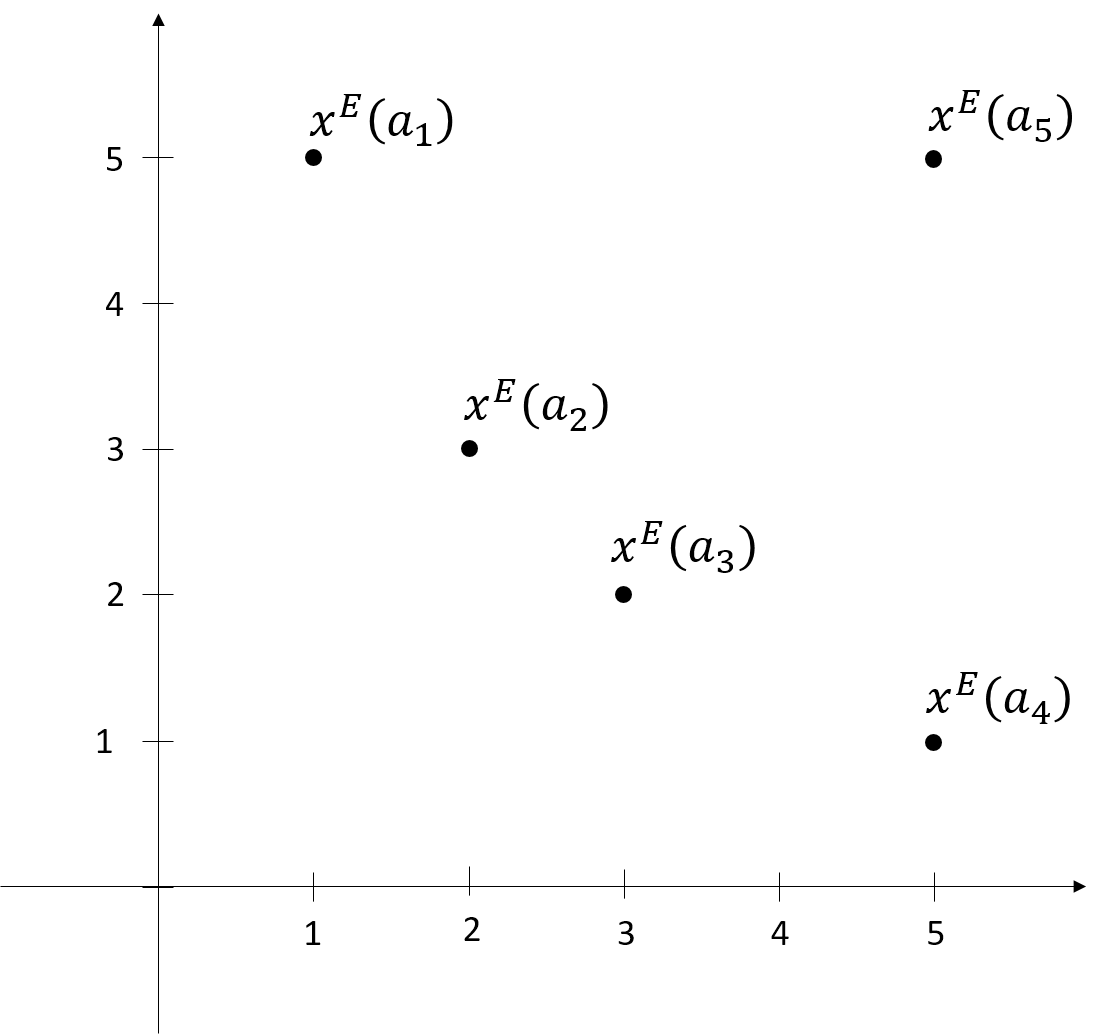}
\caption{Five alternatives with two criteria.}
\label{fig:example2c5a}
\end{figure}

\begin{example}

In figure \ref{fig:example2c5a} a simple example with two criteria and the criteria assessments of five alternatives helps to understand the new method. The goal is to order the alternatives $A = \{a_1,a_2,a_3,a_4,a_5\}$ by giving a rank to their criteria evaluations $\{x^E(a_1), x^E(a_2), x^E(a_3), x^E(a_4), x^E(a_5)\}$, where

\begin{center}
\begin{tabular}{c|cccccc}
$\omega = a_i$&$a_1$&$a_2$&$a_3$&$a_4$&$a_5$ \\ \hline
$X\of{\omega} = x^E(a_i)$&$(1,5)$&$(2,3)$&$(3,2)$&$(5,1)$&$(5,5)$\\
\end{tabular}
\end{center}

First, the $\tilde F_{X,v}(z)$ for $v^1$, $v^2$ and $v^3$ is calculated and discussed.
\begin{center}
\begin{tabular}{c|cccccc}
$a_i$					&$a_1$		&$a_2$		&$a_3$		&$a_4$		&$a_5$ \\ \hline
$x^E(a_i)$				&$(1,5)$	&$(2,3)$	&$(3,2)$	&$(5,1)$	&$(5,5)$\\ \hline
$\tilde F_{X,v^1}(x^E(a_i))$	&$0.4$		&$0.4$		&$0.6$		&$0.8$		&$1$\\
$\tilde F_{X,v^2}(x^E(a_i))$	&$0.8$		&$0.4$		&$0.4$		&$0.8$		&$1$\\
$\tilde F_{X,v^3}(x^E(a_i))$	&$0.8$		&$0.6$		&$0.4$		&$0.4$		&$1$
\end{tabular}
\end{center}
The points in figure \ref{fig:example2c5a} are symmetrically distributed around the $v^2 = (1,1)$ vector. Therefore, it is clear that for $v^2$ the criteria evaluations of $a_2$ and $a_3$, respectively $a_1$ and $a_4$ have the same rank (probability). The opposite applies to  $v^1 = (2,1)$ and $v^3 = (1,2)$, whereas for the former importance vector $x^E(a_1)$ and $x^E(a_2)$ have the lowest probability and for the latter $x^E(a_3)$ and $x^E(a_4)$ are ranked the lowest. This is simply due to the higher value assigned by $v^1$ to the horizontal axis and respectively by $v^3$ to the vertical axis.

Second, the five points in figure \ref{fig:example2c5a} are ranked with help of the cone distribution function and compared to the individual ranking of each judge.

\begin{table}[H]
	\centering
	\begin{tabular}{c|cccccc}
							&$x^E(a_1)$	&$x^E(a_2)$	&$x^E(a_3)$	&$x^E(a_4)$	&$x^E(a_5)$ \\ \hline
	$\tilde F_{X,K_A}(z)$	&$0.4$		&$0.2$		&$0.2$		&$0.4$		&$1$\\ \hline
	$\tilde F_{X,v^1}(z)$	&$0.4$		&$0.4$		&$0.6$		&$0.8$		&$1$\\
	$\tilde F_{X,v^2}(z)$	&$0.8$		&$0.4$		&$0.4$		&$0.8$		&$1$\\
	$\tilde F_{X,v^3}(z)$	&$0.8$		&$0.6$		&$0.4$		&$0.4$		&$1$
	\end{tabular}
	\caption{$\tilde F_{X,K_A}(z)$ vs. $\tilde F_{X,v}(z)$}
\end{table}

By comparing the ranking derived from the cone distribution function $\tilde F_{X,K_A}(z)$ against the ``scalarized CDFs'' $\tilde F_{X,v}(z)$, three things stand out. First, $\tilde F_{X,K_A}(z)$ assigns the lowest values. This is simply due to the infimum in its definition, where the vector $v \in K_I$ that minimizes $\tilde F_{X,v}(z)$ is chosen. Second, in all cases $x^E(a_5)$ has a rank equal to one. This is inherent in the definition of $\tilde F_{X,v}(z)$, as  $\#\cb{x_E(a_i) \mid a_i \in A, x_E(a_i) \in x_E(a_5) - H^+(v)} = m, \forall v \in K_I$. Third, for $\tilde F_{X,K_A}(z)$ the elements $x_E(a_1)$ and $x_E(a_4)$, respectively $x_E(a_2)$ and $x_E(a_3)$ have the same position, as for the judge $v^2$. The importance cone $K_I$ is generated by the vectors $v^1 = (2,1)$ and $v^3 = (1,2)$, which are symmetric around the vector $v^2 = (1,1)$. Now, if the importance cone $K_I$ is symmetric around $v^1$ as well as the criteria evaluations of the alternatives (see figure \ref{fig:example2c5a}), then some elements are ranked equally. However, if $K_I$ is spanned by $v^1$ and $v^2$, then the symmetry in the ranking is lost:
\begin{table}[H]
	\centering
	\begin{tabular}{c|cccccc}
							&$x^E(a_1)$	&$x^E(a_2)$	&$x^E(a_3)$	&$x^E(a_4)$	&$x^E(a_5)$ \\ \hline
	$\tilde F_{X,K_A}(z)$	&$0.4$		&$0.2$		&$0.4$		&$0.8$		&$1$\\ \hline
	\end{tabular}
	\caption{$K_I$ generated by $v^1$ and $v^2$.}
\end{table} 

\end{example}

\subsection{Cone quantile}
\label{subsec:conequantile}

The set-valued quantile functions as introduced in \cite{HamelKostner18JMVA} are transferred into a decision making tool. In particular, these quantiles admit to cluster the alternative's evaluations $x^E(a_i)$ based on the rank $p$ and the judge's preferences via $K_I$ and $K_A$, respectively.

Initially, the paper \cite{HamelKostner18JMVA} defines quantiles based on the scalarization of the random vector $X$ with $w \in C^+\bs\{0\}$. These quantiles categorize the values of $X$ with respect to a $w \in C^+\bs\{0\}$. Therefore, these functions can be adapted to extract elements from a set of alternatives based on the individual opinion of a judge. Here are the empirical versions of those functions adapted to a single judge decision making, where $v \in K_I\bs\{0\}$. The function $\tilde Q^-_{X, v} \colon (0,1) \to \mathcal P(\R^d)$ defined by
\begin{align*}
\tilde Q^-_{X, v}(p)	&= \cb{z \in \R^d \mid \tilde F_{X,v}(z) \geq p} = \\
						&= \cb{z \in \R^d \mid \#\cb{x^E(a_i) \mid a_i \in A, x^E(a_i) \in z - H^+(v)} \geq mp},
\end{align*} 
is called the {\em lower $v$-quantile function} of $X$, and the function $\tilde Q^+_{X, v} \colon (0,1) \to \mathcal P(\R^d)$ defined by
\begin{align*}
\tilde Q^+_{X, v}(p)	&= \cb{z \in \R^d \mid \#\cb{x^E(a_i) \mid a_i \in A, x^E(a_i) \in z - \Int H^+(v)} \leq mp},
\end{align*}
is called the {\em upper $v$-quantile function} of $X$. Both functions have convex values, in particular $\tilde Q^-_{X, v}(p)$ has upward directed (with respect to $v$) halfspaces and $\tilde Q^+_{X, v}\of{p}$ has downward directed halfspaces.

It turns out that the two sets are also useful concepts from a decision making point of view. Since, $\tilde Q^-_{X, v}(p)$ includes all $x^E(a_i)$ that for a given rank $p$ and an importance vector $v$ have a rank higher or equal to $p$, whereas $\tilde Q^+_{X, v}(p)$ includes all $x_E(a_i)$ that have a rank lower or equal to $p$. Therefore, $\tilde Q^-_{X, v}(p)$ is the set of better ranked (``good'') elements and $\tilde Q^+_{X, v}(p)$ the set of ``bad'' elements, with respect to the rank $p$ and the importance vector $v$. Moreover, the higher the $p$ the better ranked and the fewer the elements in $\tilde Q^-_{X, v}(p)$. The question is: what are good choices for $p$? This depends on the available alternatives as well as on the judges' views on the criteria.

\begin{figure}[H]
\centering
\begin{minipage}{.45\textwidth}
	\centering
	\includegraphics[width=0.9\textwidth]{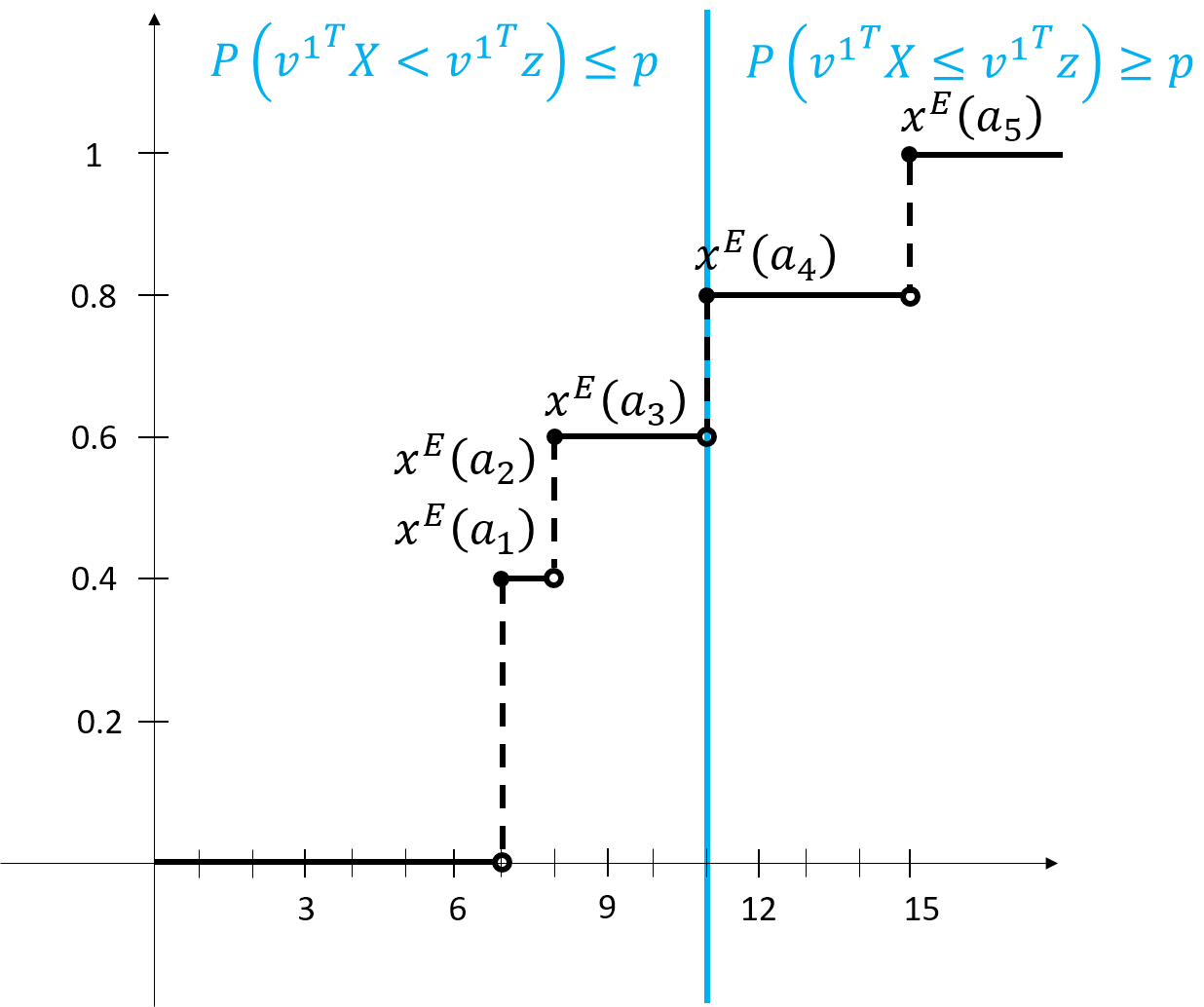}
	\caption{Graph of the CDF for ${v^1}^\top X$.}
	\label{fig:v1CDF}
\end{minipage}
\begin{minipage}{.45\textwidth}
	\centering
	\includegraphics[width=0.9\textwidth]{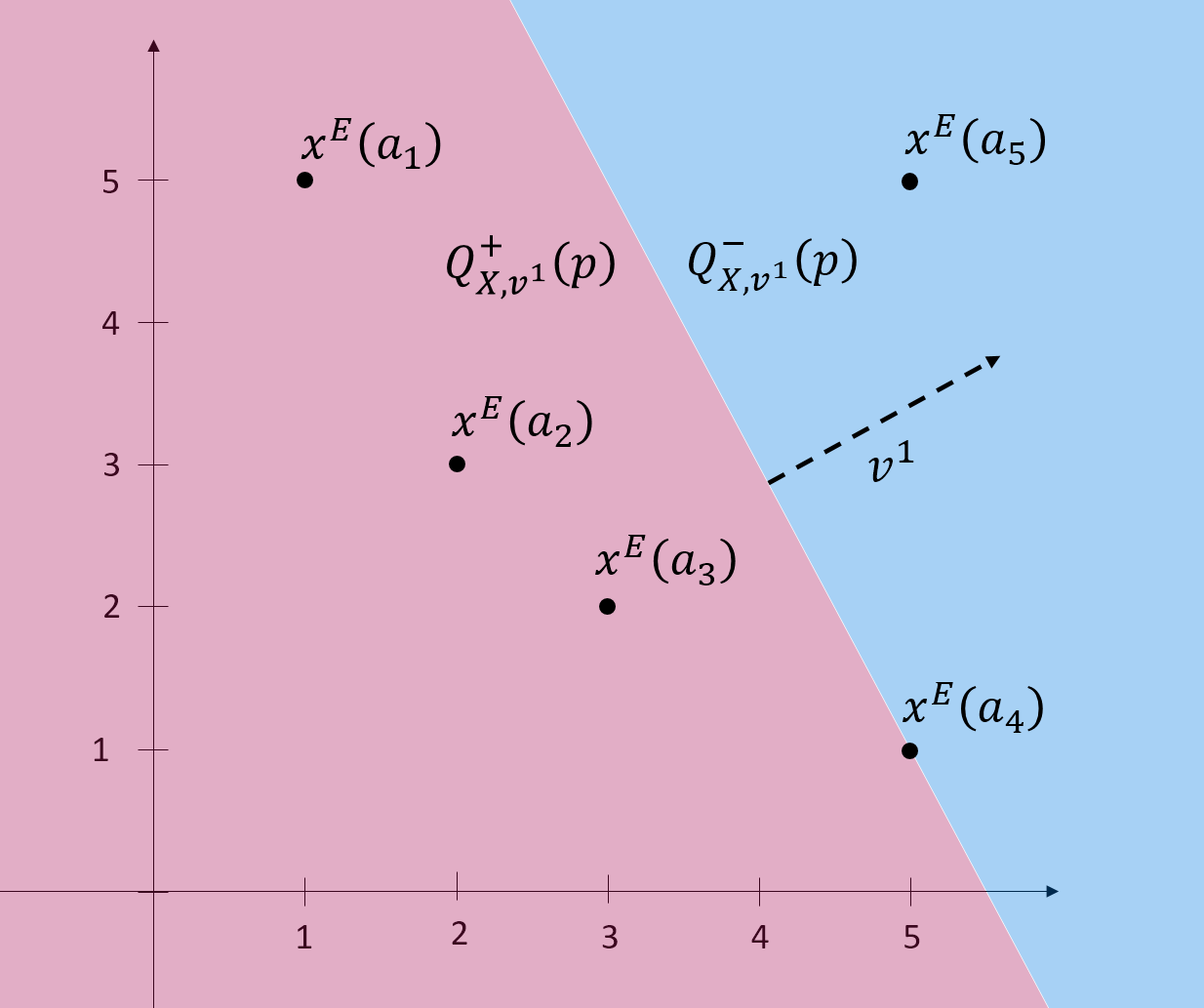}
	\caption{$\tilde Q^-_{X, v^1}(p)$ and $\tilde Q^+_{X, v^1}\of{p}$ for $v^1 = (2,1)$ and $p = 0.8$.}
	\label{fig:wQuantile}
\end{minipage}
\end{figure}

\begin{example}

Figure \ref{fig:v1CDF} and figure \ref{fig:wQuantile} illustrate the $v$-quantile functions for $X$ being the elements in \ref{fig:example2c5a}, $v^1 = (2,1)$ and $p = 0.8$. Figure \ref{fig:v1CDF} shows the graph of the cumulative distribution function for the scalarized random vector ${v^1}^\top X$, $\tilde F_{X,v^1}(z)$. Figure \ref{fig:wQuantile} has on its axes the criteria values and depicts the quantiles as halfspaces.

As shown in figure \ref{fig:v1CDF}, the ``scalar'' quantile of order $p=0.8$ for the random variable ${v^1}^\top X$ is the assessed alternative $x^E(a_4)$ scalarized by $v^1$.

From a decision making perspective one would always choose an element from $Q^-_{v_{1}^\top X}(0.8)$ over an element from $Q^+_{{v^1}^\top X}(0.8)$. This is simply due to the former quantile having higher ranked elements.

On one side, $Q^+_{{v^1}^\top X}(0.8)$ includes the elements $x^E(a_1) = (1,5)$, $x^E(a_2) = (2,3)$ and $x^E(a_3) = (3,2)$. The former two have the lowest ranks. This is simply due to their low values for the first criteria (horizontal axis) and the judge deeming the first criteria more important as the second one (vertical axis). It follows that these elements are not recommendable. The element $x^E(a_3)$ has a higher value for the criteria on the horizontal axis. However, both criteria have low values. Therefore, also the element $x^E(a_3)$ is not a good choice.

On the other side, $Q^-_{{v^1}^\top X}(p)$ includes $x^E(a_4) = (5,1)$ and $x^E(a_5) = (5,5)$, with $x^E(a_4)$ having a high value for the first criteria and $x^E(a_5)$ having the highest values for both criteria. Therefore, both elements are good options. Thus, $Q^-_{{v^1}^\top X}(p)$ can be interpreted as the set of recommendable elements.

\end{example}

The next step is to categorize a set of alternatives by complying to all judges' opinions as well as their compromises. This is done intuitively by taking the intersection of the sets $\tilde Q^-_{X, v}(p)$, and respectively $\tilde Q^+_{X, v}(p)$, over all directions $v$ of the importance cone $K_I$. This results in quantiles with the order relation $\leq_{K_A}$. As for the cone distribution function, this has the positive consequence that the sets do not have to compel to single judges but can accommodate also compromises between judges. On one hand, the \textit{lower $K_A$-quantile} $\tilde Q^-_{X, K_A} \colon (0,1) \to \mathcal P(\R^d)$ is defined as
\[
\tilde Q^-_{X,K_A}(p) = \bigcap_{v \in {K_I}\bs\{0\}} \tilde Q^-_{X, v}(p),
\]
on the other hand, the \textit{upper $K_A$-quantile} $Q^+_{X, K_A} \colon (0,1) \to \mathcal P(\R^d)$ is defined as
\[
\tilde Q^+_{X,C}\of{p} = \bigcap_{v \in {K_I}\bs\{0\}} \tilde Q^+_{X, v}\of{p}.
\]
The lower $K_A$-quantile can be also defined with respect to the cone distribution function $\tilde F_{X,K_A}(z)$:
\[
\tilde Q^-_{X,K_A}(p ) = \cb{z \in \R^d \mid \tilde F_{X,K_A}(z) \geq p}.
\]
On the one side, lower quantile $\tilde Q^-_{X,K_A}\of{p}$ is directed ``upwards'' with respect to the order relation $\leq_{K_A}$. This means that the lower quantile maps into a collection of sets, which are directed upwards with respect to the preorder generated by the cone $K_A$, refer to \cite{HamelKostner18JMVA} for further details. On the other side, the upper quantile $\tilde Q^+_{X,K_A}\of{p}$ is directed ``downwards''. This has an important implication for the decision making process. The lower quantile is the set to consider in order to extract alternatives that are ``recommended'' by the judges, whereas the upper quantile defines an area of ``non-advisable'' elements.

The potential of these quantiles as decision making tools is shown by interpreting their most relevant properties from a decision making perspective. It is enough to discuss the properties of the lower $K_A$-quantile $Q^-_{X,K_A}\of{p}$ (see \cite{HamelKostner18JMVA}), as one can easily transfer these properties into those for the upper quantiles.

\textbf{a)} For all $b \in \R^d$ and all invertible matrices $A \in \R^{d \times d}$ it holds 
\[
\forall p \in (0, 1) \colon \tilde Q^-_{AX+b, AK_A}\of{p} = A \tilde Q^-_{X, K_A}\of{p} + b.
\]
If the scale for the criteria changes, the quantile can be easily adapted for proportional variations in all criteria--$b$ or/and changes in single criteria--$A$.

\textbf{b)} If $p_1, p_2 \in (0, 1)$, $p_1 \geq p_2$, then $\tilde Q^-_{X, K_A}\of{p_1} \subseteq \tilde Q^-_{X, K_A}\of{p_2}$. By increasing the rank $p$ the quantile gets more selective and excludes more alternatives. Therefore, the higher the parameter $p$, the less and the better ranked the elements in $\tilde Q^-_{X,K_A}\of{p}$.

\textbf{c)} If $X \leq_{K_A} Y$, then $\tilde Q^-_{X, K_A}(p) \supseteq \tilde Q^-_{Y, K_A}(p)$ for all $p \in (0,1)$. $X \leq_{K_A} Y$ means that $Y$ has better--with respect to the judges' perspectives-- choices than $X$. For a given rank $p$ and importance cone $K_I$, the quantile $\tilde Q^-_{X, K_A}(p)$ of the set with the inferior choices $X$ includes the quantile $\tilde Q^-_{Y, K_A}(p)$ of the ``better'' set $Y$. This can be interpreted as $\tilde Q^-_{Y, K_A}(p)$ being more selective due to its derivation from a ``better'' set.

\textbf{d)} If $\emptyset \neq K_A^1 \subseteq K_A^2 \subset \R^d$ are two closed convex cones, then $\tilde Q^-_{X, K_A^1}(p) \subseteq \tilde Q^-_{X, K_A^2}(p)$ for all $p \in (0,1)$. As discussed above in property d of the cone distribution, a narrower acceptance cone $K_A$ implies a bigger compromise to be made, due to a wider variety in judges' opinions represented by a wider importance cone $K_I$. This effort of accommodating all the judges' opinions, is reflected in a ``smaller'' quantile, subsequently in one that includes less elements. \\

The following two examples reveal further interesting insights for the decision making process. Example \ref{QuantileClassification} illustrates the classification of the alternatives via the quantiles, whereas example \ref{QuantilesNoMatch} shows the ability of the quantile to reveal the fit between the set of alternatives and the judges' opinions.

\begin{figure}[H]
\centering
\begin{minipage}{.3\textwidth}
  \centering
  \includegraphics[width=0.95\textwidth]{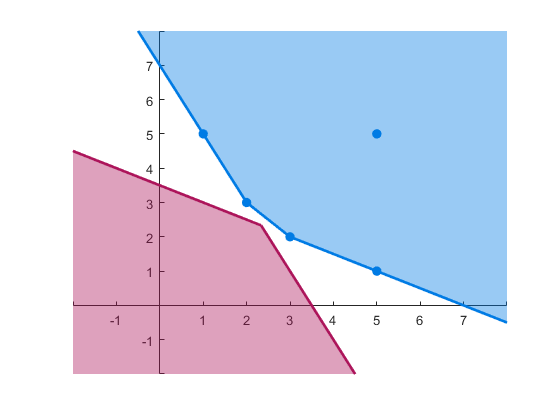}
  \caption*{$p=(0,0.2]$}
  \label{fig:ExQ1}
\end{minipage}
\begin{minipage}{.3\textwidth}
  \centering
  \includegraphics[width=0.95\textwidth]{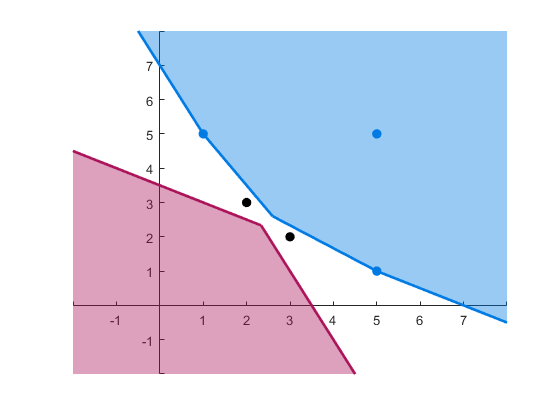}
  \caption*{$p=(0.2,0.4)$}
  \label{fig:ExQ11}
\end{minipage}
\begin{minipage}{.3\textwidth}
  \centering
  \includegraphics[width=0.95\textwidth]{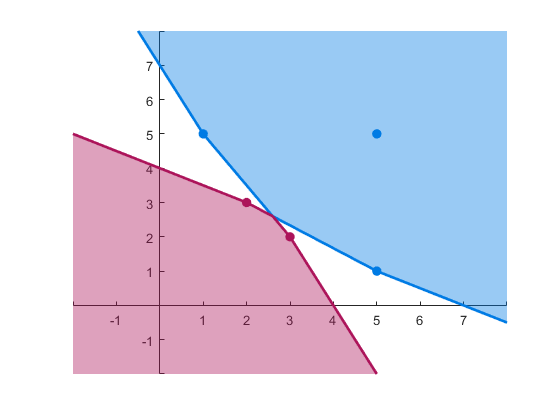}
  \caption*{$p=0.4$}
  \label{fig:ExQm}
\end{minipage}
\vskip\baselineskip
\begin{minipage}{.3\textwidth}
  \centering
  \includegraphics[width=0.95\textwidth]{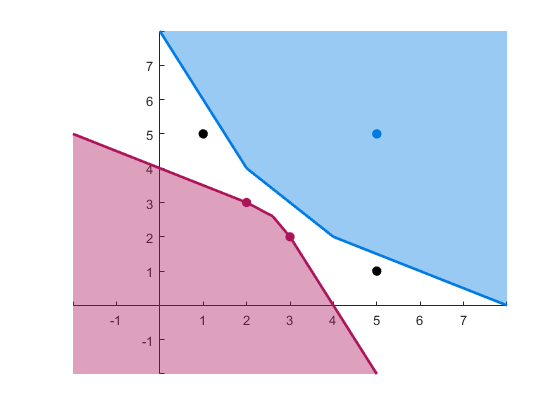}
  \caption*{$p=(0.4,0.6)$}
  \label{fig:ExQ21}
\end{minipage}
\begin{minipage}{.3\textwidth}
  \centering
  \includegraphics[width=0.95\textwidth]{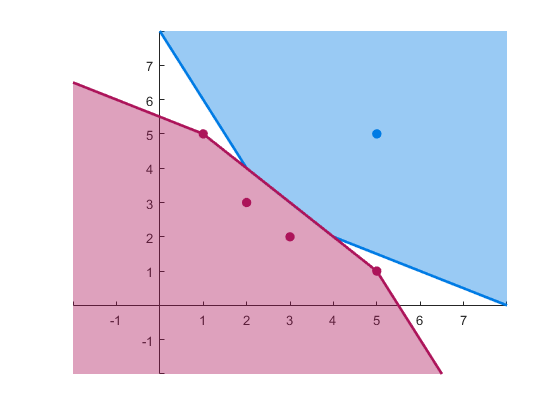}
  \caption*{$p=0.6$}
  \label{fig:ExQ3}
\end{minipage}
\begin{minipage}{.3\textwidth}
  \centering
  \includegraphics[width=0.95\textwidth]{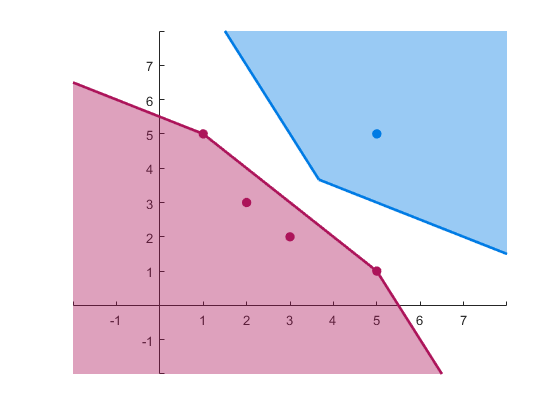}
  \caption*{$p=(0.6,0.8)$}
  \label{fig:ExQ31}
\end{minipage}

\vskip\baselineskip
\begin{minipage}{.3\textwidth}
  \centering
  \includegraphics[width=0.95\textwidth]{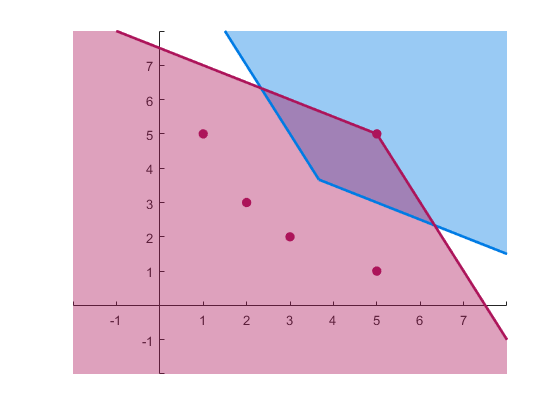}
  \caption*{$p=0.8$}
  \label{fig:ExQ4}
\end{minipage}
\begin{minipage}{.3\textwidth}
  \centering
  \includegraphics[width=0.95\textwidth]{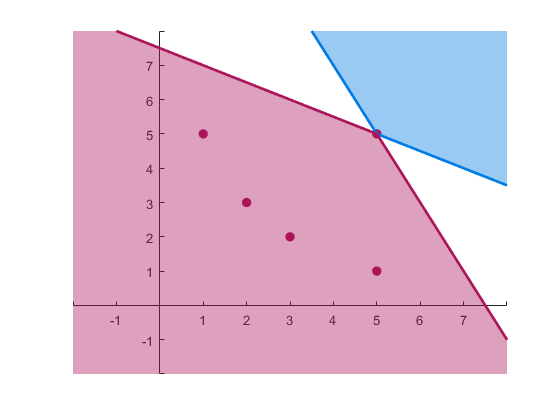}
  \caption*{$p=(0.8,0.1)$}
  \label{fig:ExQ41}
\end{minipage}
\caption{$\tilde Q^-_{X,K_A}\of{p}$ and $\tilde Q^+_{X, K_A}\of{p}$ for $p = (0,1)$ with the importance cone $K_I$ generated by the judges $v^1$ and $v^3$.}
\label{fig:ExQ}
\end{figure}

\begin{example}
\label{QuantileClassification}
Figure \ref{fig:ExQ} depicts the lower $K_A$-quantile $\tilde Q^-_{X,K_A}\of{p}$ and the upper $K_A$-quantile $\tilde Q^+_{X, K_A}\of{p}$ for different values of $p$, for the set of assessed alternatives $X$ in figure \ref{fig:example2c5a} and the importance cone $K_I$ generated by the judges $v^1$ and $v^3$.

On the one side, the intersection between $\tilde Q^-_{X,K_A}\of{p}$ and $\tilde Q^+_{X,K_A}\of{p}$ can be empty as for $p = (0,0.4)$, $p = (0.4,0.6)$ and $p = (0.6,0.8)$. On the other side, the intersection between the quantiles can be a single element as for $p = 0.4$ and $p = (0.8,1)$; a hyperplane as for $p = 0.6$; or even a set with non empty interior as for $p = 0.8$. It follows that a criteria evaluation of an alternative can be in one of the quantiles, in both quantiles or in neither of them. The assessed alternative included in only one of the quantiles can either be ``recommendable'' or ``non-recommendable''; the element included in both quantiles can be seen as ``neutral'' element, as it is neither good or bad; for the element not included in the quantiles no clear statement can be given from a decision making standpoint.
\end{example}

\begin{figure}[H]
\centering
\includegraphics[width=0.5\textwidth]{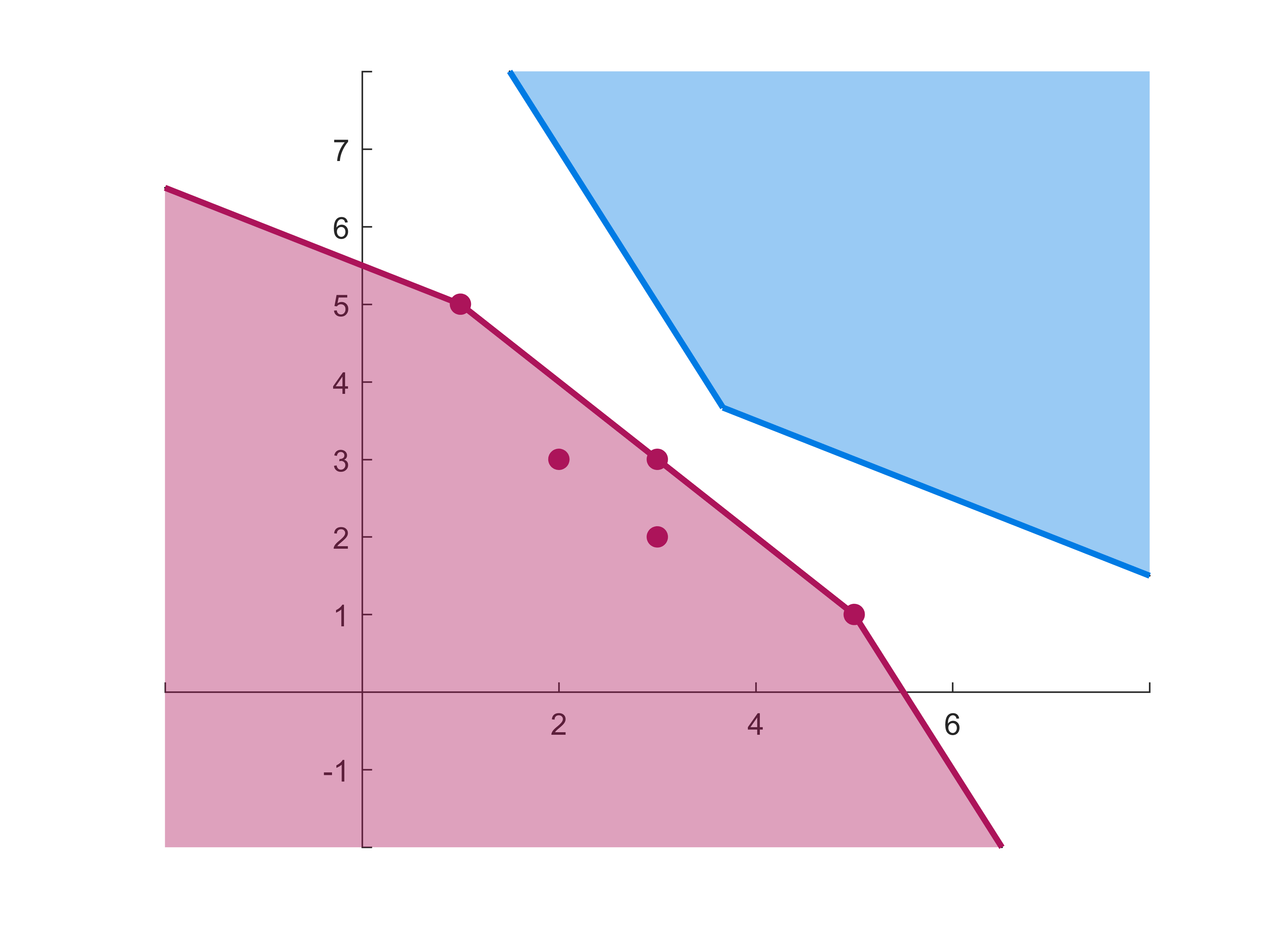}
\caption{$\tilde Q^-_{X_l,K_A}\of{p}$ and $\tilde Q^+_{X_l, K_A}\of{p}$ for $p=(0.8,1)$, where in $X_l$ is equal to $X$ with the exception that $x^E(a_5) = (3,3)$.}
\label{fig:ExQ41empty}
\end{figure}

\begin{example}
\label{QuantilesNoMatch}
The figure \ref{fig:ExQ41empty} illustrates the case when the judges' views do not really match the set of assessed alternatives $X$. The element $e$ of figure \ref{fig:example2c5a} is decreased in value for both criteria, $e = (3,3)$. This implies that $e$ is not preferable to $a$ and $d$, with regards to $\leq_{K_A}$. Consequently, the lower $C$-quantile for $p = (0.8,1)$ does not include any of the given elements in $X$. There are no elements in $X$ with a rank higher than $0.8$, which means that $X$ includes elements that only weakly satisfy the judges' views. Hence, the quantile can be used as indicator on how well the set of elements fit the judges' opinions.
\end{example}

From the discussion above, the utility of the $C$-quantiles for the multiple judge, multiple criteria decision making problem is apparent. It has the ability to derive a set of alternatives that conforms to the judges' opinions with the parameter $p$ indicating how selective the choice is. Moreover, the capability to distinguish between bad choices--upper quantile and good choices--lower quantile is very useful for analytic decision making.

\section{Conclusion}
\label{subsec:conclusion}
A novel approach in solving the multiple judge, multiple criteria decision making problem is proposed. Before a choice between alternatives can be made, those alternatives need to be ranked. On the one side, if the alternatives are valued based on a single criterion, they can be clearly ranked. On the other side, if the alternatives are valued based on multiple criteria, it gets complicated to rank them. From a mathematical standpoint, this is due to the lack of a natural ordering in higher dimensions. The set-valued quantile overcomes this issue by introducing an order relation based on convex cones. The fundamental idea in this paper is that the cones represent the judges' opinions. Based on that, the cone distribution can be implemented as a ranking function and the set-valued quantiles as functions that categorize alternatives into different sets of ``good'' and ``bad'' choices. Moreover, the paper shows that the properties of these functions ensure a reasonable decision making process. A computational procedure for the set-valued quantiles is under development. It is based on ideas from computational geometry (see \cite{RousseeuwHubert15ArX}). Due to $X$ being an empirical distribution, the method of choice would be the solution of linear vector optimization problems, which can be solved by tools closely related to \cite{LoehneWeissing16EJOR} and the references therein. A further advancement is to conceptualize a multi-criteria recommender system (see \cite{ManouselisCostopoulou2007}) based on the developments in this work. This recommender system could be designed such that it adjusts more closely to the users preferences and, therefore gives targeted and accurate suggestions.

\bibliographystyle{plain}

\end{document}